# An Actuarial Cost and Revenue Model for Helicopter Emergency Medical Services: Estimating Population-Based Coverage and Sustainability Thresholds

Robert D. Lieberthal, Ph.D.[1,2]*, Sabin Ahmed, Ph.D.[3], David M. Hechtman, B.A.[3], Lauren R. Indrisano, M.S.[4], Douglas R. Amirault, B.S.[3], Susan Haas, M.D., M.Sc.[3], Varun Saraswathula, M.P.H.[5]

[1] Thomas Jefferson University, Philadelphia, PA, USA; [2] Lieberthal & Associates, LLC, Pittsburgh, PA, USA; [3] The MITRE Corporation, McLean, VA, USA; [4] Elevance Health, Indianapolis, IN, USA; [5] Congressional Research Service, Washington, DC, USA

* Corresponding author: Robert.Lieberthal@jefferson.edu | ORCID: https://orcid.org/0000-0002-2826-5540

## Abstract

Helicopter emergency medical services (HEMS) provide rapid access to critical care but are costly to operate and difficult to sustain financially. A clear understanding of these costs is essential for evaluating the feasibility and design of population-based funding or policy strategies. Can an actuarial model combining cost structure and reimbursement data estimate the per-member-per-month (PMPM) cost of sustaining a HEMS base and the annual number of transports needed for financial viability? The term 'actuarial' here refers to population-based cost estimation methods drawn from insurance pricing practice (PMPM rate development), rather than a comprehensive actuarial risk model. We developed a two-part model: (1) a cost framework capturing capital and operating expenses (e.g., aircraft, equipment, labor, facilities), and (2) an actuarial revenue model using healthcare encounter data and payer reimbursement rates. The model was applied to a commercially insured Massachusetts population (3.9M lives), using provider charge data and Medicare fee schedules. We analyzed breakeven transport volumes under varying reimbursement and labor cost assumptions, including sensitivity scenarios. Under optimistic assumptions (full charge realization, minimal overhead), breakeven is reached with approximately 90 annual transports. More realistic scenarios incorporating commercial reimbursement at 50% of charges and full 24/7 staffing require 184 transports. If labor costs are doubled or Medicare rates are used exclusively, breakeven thresholds exceed 1,000 transports per year. A Monte Carlo simulation (10,000 iterations) confirmed the robustness of these thresholds: the median simulated breakeven was 190 transports under commercial reimbursement closely matching the deterministic base case. The 90th percentile reached 304 (commercial) and 1,066 (Medicare) annual transports. HEMS programs are highly sensitive to labor costs and payer reimbursement levels. Sustainable operation requires more transport volume than previously estimated, especially when reimbursement is constrained, or staffing costs increase. This model provides a transparent, replicable tool to inform financial planning, policy evaluation, and payer negotiations for air medical services.

# 1. Introduction

## *1.1. Background*

Helicopter Emergency Medical Services (HEMS) "...are used to transport patients via helicopter or airplane in critical and life-threatening situations", often bridging the gap between rural or remote locations and advanced medical care. [1] While their clinical value is well recognized, HEMS operations are also expensive and complex, raising important questions about financial sustainability. In recent years, increased public and policy attention has focused on the costs of HEMS, with patients, providers, and insurers grappling with how to support these services in a financially responsible manner. [1]

The clinical value of HEMS has been documented across a substantial body of literature. A systematic review by Taylor et al. identified 15 economic evaluations of HEMS programs, finding annual program costs ranging from $115,777 to $5.6 million and noting wide variation in cost-effectiveness across clinical settings and patient populations.[2] Galvagno et al. demonstrated a 16% survival benefit associated with helicopter versus ground transport in a landmark U.S. analysis of more than 220,000 major trauma patients.[3] Delgado et al. were among the first to quantify the cost-effectiveness threshold in the U.S. context, estimating that HEMS must reduce mortality by at least 15% relative to ground transport to achieve a cost per quality-adjusted life-year (QALY) below $100,000.[4] More recently, a 2025 systematic review and meta-analysis of 77 studies encompassing 2.6 million patients found that the survival benefit of HEMS is heterogeneous across patient populations and settings, while the

incremental net benefit reached approximately €980,000 per QALY in cost-effectiveness analyses.[5] For patients with severe traumatic brain injury (Glasgow Coma Score <9), HEMS scene response has been specifically associated with improved survival (OR 1.37, 95% CI 1.23–1.53).[6] A scoping review of 2023 outcomes literature catalogued 18 new HEMS outcomes studies, with 78% focused on trauma, underscoring continued research interest in air medical effectiveness.[7]

Despite this evidence base, financial sustainability remains a persistent challenge for HEMS operators, particularly in the United States. Air ambulance providers rely primarily on commercial insurance reimbursement; Medicare and Medicaid rates, which apply to an estimated 70% of transported patients alongside the uninsured, have not been substantially updated since 2002 and cover only approximately 50% of current operating costs.[1] The No Surprises Act, effective January 2022, further constrains revenue by prohibiting balance billing of privately insured patients for emergency air transport.[8] International modeling studies have documented similar structural cost pressures: a German scenario analysis estimated total annual costs of €1.7 million for a single 12-hour base, achieving breakeven at approximately 728 annual missions under an engine-runtime reimbursement model.[9] A Finnish cost-effectiveness analysis found that fixed costs account for approximately 93% of total HEMS expenditures, a structural feature with direct implications for the volume-dependent breakeven thresholds that are the focus of the present study.[10]

Despite the prominence of HEMS in emergency response systems, there is limited research quantifying the financial requirements of operating a base. Much of the existing literature has focused on clinical outcomes, cost-effectiveness comparisons with ground transport, or reimbursement policy. However, understanding the full cost structure, including fixed capital investments, staffing, aviation operations, and medical equipment, is essential for evaluating the financial viability of HEMS, particularly in population-based funding contexts.

### *1.2. Motivation*

This paper introduces an actuarial modeling framework to assess the per-member-per-month (PMPM) cost of sustaining a rotary-wing HEMS base. The term "actuarial" refers to the estimation of expected costs across a defined population and time period, a common method in insurance pricing and health system budgeting. [11] By linking cost inputs with population-level utilization and reimbursement data, this approach offers a structured way to evaluate how many transports, and how much financial support are needed to sustain a HEMS base.

Prior research has examined both the clinical value and financial structure of HEMS from multiple perspectives. Ringburg et al. synthesized the literature on lives saved by air medical programs.[12] International scenario-based cost models have characterized the fixed-cost-dominant structure of HEMS operations in Germany [9] and Finland, [10] and Delgado et al. established a cost-effectiveness threshold framework for U.S. scene transport. [4] However, none of these frameworks has been translated into the U.S. commercial insurance context, and no prior study has developed actuarially sound per-member-per-month (PMPM) estimates directly applicable to health plan pricing, global budgeting, or subsidy program design.

## 2. Materials and Methods

This study uses a structured, two-part theoretical model to examine the financial viability of a rotary-wing HEMS base. The model consists of: (1) a cost structure framework to capture fixed and variable expenses and (2) an actuarial reimbursement framework to estimate revenue under a defined population base. The integration of these two components enables estimation of the number of patient transports necessary to cover total costs under different funding and utilization scenarios.

*2.1. Theoretical Cost Model*

We conceptualize the total cost of operating a single HEMS base as the sum of capital expenditures and ongoing operating expenses:

- **Capital Expenditures (CapEx):** These include one-time investments such as aircraft acquisition and medical equipment. Aircraft costs are assumed to be financed and amortized over a multi-year period.
- **Operating Expenses (OpEx):** These include recurring costs such as personnel, fuel, insurance, maintenance, regulatory compliance, facility operations, and training.

The total annual cost of maintaining a HEMS base can therefore be expressed as:

$$TC = Aircraft\ Financing\ + Operating\ Costs\ +\ Annualized\ Medical\ Equipment\ Costs\ +\ Labor\ Costs + Facility\ Costs$$

*Equation 1: Total costs of an air ambulance base with a single aircraft*

Theoretical Actuarial (Reimbursement) Model

The revenue model is based on actuarial principles, using expected reimbursement per transport multiplied by the number of transports. Reimbursement values may be based on public fee schedules (e.g., Medicare) or private claims data; other payment models such as bundled or fixed pricing are beyond the scope of this study. Reimbursement for HEMS is typically composed of:

- *A base transport fee* $R_b$
- *A mileage-based fee* $R_m$

Expected annual total revenue is then:

$$Total\ Revenue = N \times (R_b + R_m)$$

*Equation 2: Total reimbursement revenue for air ambulance services*

Where $N$ is the number of annual patient transports. Reimbursement values may be based on charge data, average realization rates (e.g., 50%), out-of-network payment rates, or standardized government fee schedules.

We define the breakeven number of transports $N^*$ as the value for which total revenue equals total cost:

$$N^* = \frac{TC}{R_b + R_m}$$

*Equation 3: Breakeven calculation*

This relationship forms the basis for determining the minimum patient volume needed for financial viability under different assumptions.

In the following section, we apply this theoretical framework to real-world data sources, including healthcare encounter databases, salary benchmarks, and aviation industry cost references, to produce quantitative estimates for each model component.

### *2.2. AI Disclosure*

The authors used AI-assisted copy editing (Claude.ai) to improve readability and style, including grammar, spelling, punctuation, and tone. The authors also used Claude.ai to produce a draft version of the Monte Carlo simulation in the Excel spreadsheet and to organize responses to reviewers. No generative content was autonomously created. The authors reviewed and edited all content and take full responsibility for the final version of the publication.

## 3. Data

Data for the model is shown in Table 1 below. We also describe the derivation of all the data used to populate the model in the following sections.

| Category | Item | Frequency | Cost ($) | Notes/References |
|---|---|---|---|---|
| **Aircraft Costs** | Purchase | One time | 8,700,000 | EC-145 helicopter |
| **Aircraft Costs** | Loan payment | Annual | 1,568,359 | 6-year loan at 3.25% |
| **Aircraft Costs** | Operating Budget | Annual | 795,810 | Includes cost of a pilot |
| **Medical Equipment** | Capital goods | One time | 25,510* | Based on equipment needed for an ambulance including automatic external defibrillator |
| **Medical Equipment** | Equipment | Per patient | 902 | Basic life support (BLS) equipment and interventions |
| **Personnel Costs** | Pilot | Annual | 110,759** | Job 53-2012; Includes wages and salary and supplements (benefits) |
| **Personnel Costs** | EMT | Annual | 46,271 | Job 29-2041; Includes wages and salary and supplements (benefits) |
| **Personnel Costs** | Critical care nurse | Annual | 92,792 | Job 29-1141; Includes wages and salary and supplements (benefits) |

| **Personnel Costs** | Medical director | Annual | 345,970 | Job 29-1228; Includes wages and salary and supplements (benefits) |
|---|---|---|---|---|
| *** This figure includes costs gathered from sources from ground ambulances where air ambulance specific equipment was not available.** | | | | |
| **** One pilot is included in the "Operating Budget" line. This data is used to add additional FTE pilots in the Results section.** | | | | |

*Table 1: Air base cost data*

### *3.1. Aircraft costs*

Aircraft-related expenditures form the largest single capital and operational expense for a rotary-wing HEMS program. For this model, we assumed the acquisition of an EC145 twin-engine helicopter, a commonly used model in air medical transport, with a purchase cost of $8.7 million. [13] The fixed and variable costs of aircraft operations were obtained from general aircraft cost calculator resources (AircraftCostCalculator.com). Cost categories captured include fuel, insurance, maintenance, avionics, and inspections, and we used mid-range public estimates.

To reflect real-world financing, we assumed this cost is paid over a six-year loan period based on the IRS depreciation schedule for aircraft to estimate the working lifetime for a helicopter at an interest rate of 3.25%, consistent with AAA municipal bond and healthcare loan terms.[14] This results in an annualized loan payment that becomes part of the fixed annual cost. Prime and AAA municipal bond loan rates were used to estimate the interest rate for an aircraft purchase over time. [15]

Operating costs for the aircraft include fuel, insurance, scheduled and unscheduled maintenance, avionics, and inspections. These costs were drawn from publicly available aviation cost calculators and industry sources and are estimated at approximately $795,810 per year. [16]

The assumptions reflect mid-range cost estimates for hospital-owned programs and exclude rural cost multipliers or vendor-specific bundled services. All aircraft costs are presented separately from other categories to allow flexible use of the model in both capital and lease arrangements.

### *3.2. Medical equipment*

Medical equipment costs include the essential tools and supplies required to deliver emergency care in-flight. These range from basic monitoring and resuscitation tools to specialized devices suitable for airborne operations. We used a hybrid approach to estimate both initial capital and per-transport equipment expenses.

Initial capital equipment was estimated at $25,510, derived from the costs of furnishing an ambulance with standard life-support and monitoring gear, including defibrillators, suction units, infusion pumps, and ventilators. This estimate was adapted from ground ambulance procurement sources and adjusted for aviation-specific needs.

In addition to capital equipment, consumables and durable items used on a per-transport basis were estimated at $902 per patient. These costs include oxygen, medications, linens, disposables, and other expendables typically required during transport.

As with other cost elements, these figures represent mid-range national benchmarks and may vary by program protocols, medical director preferences, and supply chain arrangements. The model allows for equipment costs to be amortized over an appropriate useful life or reflected per-transport, depending on analytical goals.

### *3.3. Labor costs*

Labor represents a substantial portion of a HEMS program's annual operating expenses due to the need for continuous, 24/7 staffing. Salaries include both direct wages and supplements to wages (e.g., benefits and payroll taxes), sourced from the IMPLAN economic input-output model. [17] These assumptions represent typical compensation for a hospital-based HEMS program in the United States and may differ for vendor-operated or rural models.

Several industry experts and anonymous reviewers discussed the likelihood that the numbers in Table 1 underestimate the actual labor costs associated with these positions. We discuss a sensitivity analysis below where we double all the labor costs in order to determine how our model results vary in geographies and industries where compensation is substantially higher. [18]

### *3.4. Healthcare encounters*

Healthcare encounters data was obtained from the Clarivate (previously Decision Resources Group or DRG) Real-World Data set. We assessed population size and utilization for the state of Massachusetts. The data set covers approximately 3.9 million patients within the state. [19] The data is collected through an aggregation of claims and electronic health records from healthcare payment clearinghouses and other sources. This means that the encounters in the data represent actual healthcare episodes;

however, this is not an “all payer claims database” where the entire population is represented.

Current procedural terminology (CPT) codes A0431 and A0436 were used for estimating the cost of a one-way, rotary wing ambulance service and rotary wing mileage, respectively. These codes are published by the American Medical Association and used by healthcare payers and insurance such as the Centers for Medicare & Medicaid Services (CMS) as well as private health insurers. “The Ambulance Fee Schedule is a national fee schedule for ambulance services furnished as a benefit under Medicare Part B. The fee schedule applies to all ambulance services, including volunteer, municipal, private, independent, and institutional providers, hospitals, critical access hospitals (except when it is the only ambulance service within 35 miles), and skilled nursing facilities.” [20] A0431 is a claim for the pickup by the air ambulance itself, and A0436 is used to charge a variable amount for the miles between the pickup point and destination (typically a hospital). Reimbursement code data is shown in Table 2 below. Note that commercial plans typically reimburse 40–60% of billed charges for out-of-network emergency services; 50% is a commonly cited midpoint consistent with Turrini et al. [1].

| Code | 2022 | | 2023 | | 2024 | | 2025 | |
|---|---|---|---|---|---|---|---|---|
| | Base | Per Mile | Base | Per Mile | Base | Per Mile | Base | Per Mile |
| **A0430, A0453** | $3,170.86 | $21.85 | $3,370.14 | $23.22 | $3,570.59 | $24.59 | $3,672.32 | $25.28 |
| **A0431, A0436** | $3,686.61 | $32.41 | $3,934.18 | $34.47 | $4,181.75 | $36.53 | $4,303.99 | $37.60 |

*Table 2: Medicare Reimbursement Data for Air Ambulances*

## 4. Results

### *4.1. Utilization and Actuarial Pricing*

We present the utilization data from the DRG database for air ambulances in Table 3.

| CPT code | Meaning | Count | Sum | Average |
|---|---|---|---|---|
| **A0430** | Fixed wing (base) | 65 | $947,033.85 | $14,569.75 |
| **A0431** | Rotary (base) | 705 | $22,824,565.18 | $32,375.27 |
| **A0435** | Fixed wing (mileage) | 68 | $2,742,615.86 | $40,332.59 |
| **A0436** | Rotary (mileage) | 700 | $8,122,452.40 | $11,603.50 |

*Table 3: Air ambulance utilization in DRG data*

From this data, we estimated an average rotary-wing transport charge of $43,979, composed of a $32,375 base charge and an $11,604 mileage charge (averaging 50 miles per transport).

Given that the exposed population here is known to be approximately 3.9 million (3,879,778) it is also possible to develop an actuarial rate for these services on a managed care (per member, per month) basis. We show the rate development for this population in Table 4. Once a rate is developed in column (e), that can be applied across an insured population to build this service into a health insurance premium rate.

| | Annual utilization per 1,000 | Average charge | Gross charged costs per member per month (PMPM) | Cost sharing | Net charged cost PMPM |
|---|---|---|---|---|---|
| **HCPCS** | (a) | (b) | (c) | (d) | (e) |
| **A0430** | 0.017 | $14,569.75 | $0.02 | $500 | $0.02 |
| **A0431** | 0.182 | $32,375.27 | $0.49 | $1,000 | $0.48 |
| **A0453** | 0.018 | $40,332.59 | $0.06 | N/A | $0.06 |

| A0436 | 0.180 | $11,603.50 | $0.17 | N/A | $0.17 |
|---|---|---|---|---|---|
| **Sum (All codes)** | | | $0.74 | | $0.73 |
| **Sum (Rotary wing only)** | | | $0.66 | | $0.65 |

*Table 4: Actuarial pricing of air ambulance services (notional)*

The observed utilization rate for rotary-wing services was 0.362 per 1,000 lives annually, translating to a gross PMPM charge of $0.66. Assuming a notional patient cost share of $1,000 per episode, the net PMPM cost to payers was $0.65.

Given the amounts that are given in each row, the total per member per month (PMPM) budgeted cost for air ambulance services for an insurer is $0.73, which is the sum of the net costs for each amount charged, or $0.74 PMPM for the total amount (insurer and member).

### 4.2. Medicare Pricing

We also determined the cost of these CPT codes based on the Medicare fee schedule. This shows how the numbers in Table 4 would change with the use of Medicare fee schedule prices for 2023.

| | Annual utilization per 1,000 | Medicare Reimbursement | Gross charged costs per member per month (PMPM) | Cost sharing | Net charged cost PMPM |
|---|---|---|---|---|---|
| **HCPCS** | **(a)** | **(b)** | **(c)** | **(d)** | **(e)** |
| A0431 | 0.182 | $3,934. 18 | $0.06 | N/A | $0.06 |
| A0436 | 0.180 | $1,723.50* | $0.03 | N/A | $0.03 |
| Total | N/A | 5,657.68 | $0.09 | N/A | $0.09 |
| Notes:<br>* Assumes a trip of 50 miles | | | | | |

*Table 5: Pricing based on 2023 Medicare Reimbursement*

In 2023, the national Medicare base rate for rotary-wing transport was $3,934, with a mileage reimbursement of $34.47 per mile. For a 50-mile average transport, this yields an average payment of $5,658 per case. When applied to the observed utilization rate, the resulting PMPM figure is $0.09, which is substantially lower than the PMPM based on provider charges. This difference illustrates the wide gap between nominal charges and standardized public payer reimbursement levels.

### *4.3. Cost Analysis*

Annual costs of maintaining a single rotary-wing HEMS base were estimated by combining amortized capital costs, labor, and operations. A six-year loan for aircraft purchase ($8.7M) and equipment amortization were included. To ensure safe and consistent coverage, most programs require approximately 4.5 to 5.0 full-time equivalents (FTEs) per role to cover all shifts, training, vacations, and administrative time. Labor costs were based on national wage data including benefits.

Total annual fixed costs were estimated at approximately $3.85 million, excluding variable patient-level equipment costs. Variable costs added approximately $902 per patient. These values served as the basis for determining transport volume thresholds needed to meet financial breakeven under various reimbursement assumptions. In particular, we note that fixed costs represent approximately 95% of total costs at the base-case breakeven volume of 184 transports shown below.

| Category | Item | Frequency | Cost ($) | Notes/References |
|---|---|---|---|---|
| **Personnel Costs** | Pilot | Annual | 443,036 | 5.0 FTE (4 in addition to Operating Budget) |
| **Personnel Costs** | EMT | Annual | 231,355 | 5.0 FTE |

| **Personnel Costs** | Critical care nurse | Annual | 463,960 | 5.0 FTE |
|---|---|---|---|---|
| **Personnel Costs** | Medical director | Annual | 345,970 | 1.0FTE |
| **Fixed annual costs** | | | 3,874,000 | Includes other fixed costs from Table 1 |
| **Additional costs** | | | | |
| **50 patients** | | | 45,100 | |
| **100 patients** | | | 90,200 | |
| **250 patients** | | | 225,500 | |
| **1000 patients** | | | 902,000 | |

*Table 6: Air base cost results*

### *4.4. Breakeven Analysis*

By integrating the cost and revenue models, we calculated the breakeven number of annual transports, or the number required for revenue to match costs, under three scenarios:

- **Optimistic scenario**: Charges are reimbursed in full (100% realization)
- **Base-case scenario**: 50% realization of charges, reflecting typical commercial plan payments
- **Conservative scenario**: Reimbursement set at CMS fee schedule rates only

By integrating these revenue models with the cost model described earlier, we calculated the breakeven number of transports required for a HEMS base to remain financially sustainable under each scenario. The breakeven point is defined as the number of annual transports that satisfies Equation 3.

The table below summarizes the breakeven transport volumes across the three modeled reimbursement scenarios:

| Reimbursement Scenario | Labor Cost Scenario | Assumed Realization | Reimbursement | Patients | Revenue | Costs |
|---|---|---|---|---|---|---|
| **Optimistic (100% charges)** | Base | 100% | 43,979.00 | 90 | 3,958,110 | 3,955,180 |
| | Sensitivity | | | 127 | 5,585,333 | 5,583,634 |
| **Base-case (Commercial Avg)** | Base | 50% | 21,989.50 | 184 | 4,046,068 | 4,039,968 |
| | Sensitivity | | | 260 | 5,717,270 | 5,703,600 |
| **Conservative (CMS rates)** | Base | Medicare rates | 5,657.68 | 815 | 4,611,009 | 4,609,130 |
| | Sensitivity | | | 1,150 | 6,506,332 | 6,506,380 |

*Table 7: Breakeven analysis*

These results show that, although PMPM costs may appear modest from a population-wide view, the sustainability of a HEMS base depends heavily on both payer reimbursement levels and transport volume.

### *4.5. Sensitivity Analysis*

To test the robustness of the model, we simulated a high-cost labor market by doubling all labor costs. Under this assumption, annual fixed costs increased to over $5.6 million. Breakeven analysis revealed that:

- Under full charge reimbursement, breakeven volume rose from 90 to 127 transports.
- Under 50% realization, breakeven volume increased to 260 transports.
- Under Medicare-only reimbursement, breakeven exceeded 1,000 transports annually.

This sensitivity analysis highlights labor as the dominant cost driver in air medical operations and underscores the difficulty of sustaining service in low-volume or low-reimbursement environments.

### *4.6. Probabilistic Sensitivity Analysis*

To complement the deterministic sensitivity analysis and characterize the uncertainty inherent in individual model parameters, we conducted a Monte Carlo simulation of breakeven transport volumes. Ten thousand iterations were executed using the RunMonteCarlo macro embedded in the supplementary Excel workbook available as part of the supplementary data. In each iteration, parameter values were drawn from the following distributions:

- Realization rate: Normal(0.50, 0.167), clipped to [0.05, 0.99], approximating a Beta(4,4) distribution centered on the base-case commercial realization assumption
- Miles per transport: LogNormal($\mu = 3.85$, $\sigma = 0.385$), reflecting observed variation in transport distance
- Labor cost multiplier: LogNormal($\mu = -0.044$, $\sigma = 0.294$), reflecting regional wage variation
- Aircraft operations cost multiplier: LogNormal($\mu = -0.011$, $\sigma = 0.149$), reflecting maintenance cost uncertainty
- Variable cost multiplier: LogNormal($\mu = -0.020$, $\sigma = 0.198$), reflecting supply cost variation

For each iteration the breakeven transport volume was computed separately under commercial reimbursement and Medicare fee schedule assumptions. Results are presented in Table 8 (breakeven distribution by percentile) and Table 9 (probability of achieving financial viability at selected annual transport volumes).

| Percentile | Commercial Reimb. (transports/yr) | Medicare Fee Sched. (transports/yr) | Comm. Deterministic (Table 7) | Medic. Deterministic (Table 7) |
|---|---|---|---|---|
| **5th** | 111 | 612 | — | — |
| **10th** | 124 | 660 | — | — |
| **25th** | 150 | 743 | — | — |
| **50th (median)** | 190 | 847 | 184 | 815 |
| **75th** | 253 | 958 | — | — |
| **90th** | 304 | 1,066 | — | — |
| **95th** | 461 | 1,134 | — | — |
| **Mean** | 248 | 858 | — | — |
| **Std. Dev.** | 160 | 160 | — | — |

*Table 8: Breakeven transport volume distribution by percentile — Monte Carlo Simulation (10,000 Iterations)*

| Annual Transports | P(Viable) Commercial | P(Viable) Medicare | Interpretation |
|---|---|---|---|
| 100 | 46.2% | 0.0% | Very unlikely under either reimbursement scenario |
| 184 | 76.4% | 0.0% | ~76% probability under commercial; near-zero under Medicare |
| 260 | 95.5% | 0.4% | ~95% probability under commercial; rare under Medicare (<1%) |
| 500 | 98.5% | 38.4% | Near-certain commercial; ~38% probability under Medicare |
| 800 | 99.0% | 82.5% | Near-certain commercial; ~83% probability under Medicare |
| 1,000 | 99.4% | 90.0% | Near-certain under both scenarios |
| 1,500 | 99.9% | 99.9% | Near-certain under both scenarios (>99%) |

*Table 9: Probability of achieving financial viability at selected annual transport volumes (10,000 Iterations)*

The median simulated breakeven under commercial reimbursement (190 transports, P50) closely matched the deterministic base-case estimate of 184 transports, validating the model's internal consistency. The 90th-percentile commercial breakeven reached 304 transports annually, while the corresponding Medicare threshold exceeded 1,066 transports. The probability-of-viability analysis (Table 9) shows that at 260 annual transports, commercial reimbursement yields viability with approximately 95.5% probability, while Medicare viability remains below 1%. At 800 transports, commercial viability exceeds 99% and Medicare viability reaches approximately 82.5%. These results confirm that the deterministic base case represents a central estimate and that parameter uncertainty does not materially alter the paper's core conclusions.

Figure 1 presents histograms of the simulated breakeven distributions for commercial and Medicare reimbursement scenarios, with deterministic base-case values and key percentiles (P10, P50, P90) overlaid.

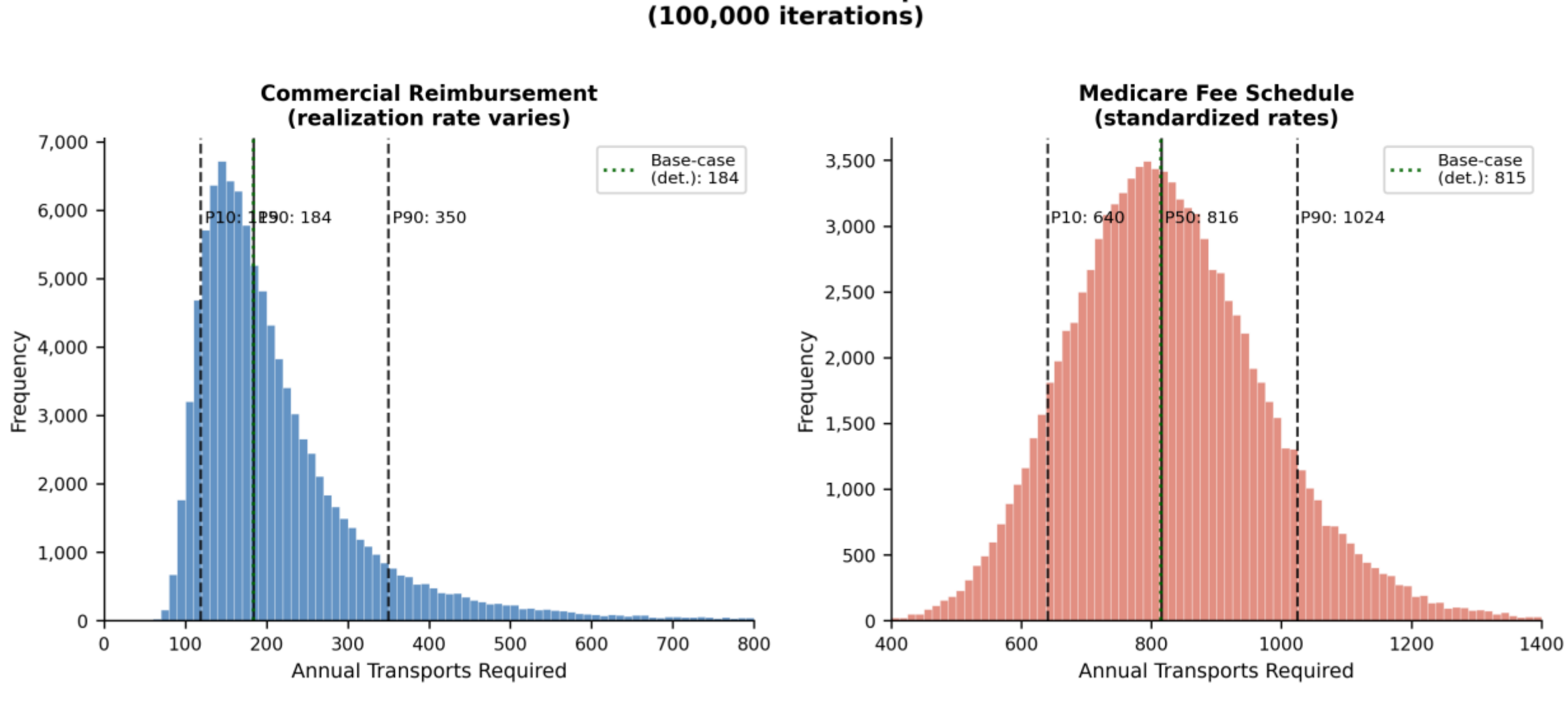

*Figure 1: Histograms of the simulated breakeven distributions for commercial and Medicare reimbursement scenarios*

## 5. Discussion

This analysis presents an actuarial framework for estimating the financial sustainability of helicopter emergency medical services (HEMS), linking fixed and variable cost structures with reimbursement scenarios to evaluate transport volumes required to support operations. The findings demonstrate that financial viability is highly sensitive to core assumptions, particularly labor costs and reimbursement rates. While early modeling under optimistic assumptions suggested breakeven could be achieved with as few as 90 transports annually, this threshold rose sharply to 260-800+ transports under more realistic labor and reimbursement assumptions. When labor costs are doubled or when only Medicare reimbursement rates are used, breakeven volumes exceed 1,000 transports per year.

Recognizing about generalizability, we emphasize that our modeled population (3.9M) reflects only the commercially insured segment of Massachusetts and excludes Medicaid, Medicare, uninsured, and other groups. While this limits extrapolation to the full state population (approximately 7.1M), this commercially insured subset was chosen for consistency with actuarial pricing methods used in health insurance rate development. We discuss this limitation explicitly in the interpretation of our per-member-per-month estimates and sensitivity scenarios.

These results underscore the need for careful financial planning, particularly in regions with limited patient volume or constrained payer reimbursement. The sensitivity analysis highlights labor costs as the primary cost driver, emphasizing the importance of staffing model design, regional wage benchmarks, and personnel flexibility in sustaining air medical programs.

In practice, HEMS programs compensate for Medicare/Medicaid shortfalls through cross-subsidization from commercial contracts, hospital or health system operating subsidies, municipal grants, state appropriations, and direct-to-consumer subscription models. The No Surprises Act has further constrained revenue by limiting balance billing, increasing pressure on network participation and public subsidy. Our model quantifies breakeven requirements under each reimbursement scenario but does not model these compensatory mechanisms, which are important directions for future work.

This model offers several contributions to the literature: it introduces a transparent method for translating cost and utilization data into per-member-per-month (PMPM) actuarial pricing; it identifies breakeven thresholds under a variety of payer conditions; and it allows for modular adaptation to specific settings or service models. The thresholds presented here should be interpreted as illustrative estimates for strategic planning and policy evaluation, not as precise operational forecasts; actual program economics will vary based on local labor markets, aircraft type, payer mix, and cost categories included.

There are several important directions for future research. First, this model could be extended to include multiple payers such as Medicare, Medicaid, commercial plans, and uninsured patients, each with their own reimbursement structures and realization rates. This would allow a more granular understanding of how payer mix influences financial viability.

Second, future studies should explicitly differentiate between urban and rural operating environments. Rural bases may face higher fixed costs due to crew housing, travel logistics, or lower economies of scale, and may benefit from policy subsidies or geographic rate adjustments. Urban bases, by contrast, may experience higher demand density but also face more stringent competition and cost-of-living pressures.

Third, the current model focuses on helicopter (rotary-wing) emergency transport. However, the majority of air ambulance transports in many systems, particularly interfacility transfers, involve fixed-wing aircraft. Future extensions should adapt the cost and revenue framework to fixed-wing operations, accounting for differences in aircraft costs, crew composition, and longer transport distances.

Finally, future research could validate and refine these model assumptions using real-world operational data from a diverse set of air medical programs, including not-for-profit hospital-based and commercial vendor models. Linking actuarial modeling with observed performance metrics, such as realized revenue, transport volume, and patient acuity, would strengthen its utility for both policy development and organizational planning.

## 6. Conclusions

This study presents a transparent, replicable actuarial framework for estimating the financial sustainability of helicopter emergency medical services. HEMS breakeven transport volumes range from approximately 90 under optimistic reimbursement assumptions to over 1,000 annually when labor costs are elevated or only Medicare rates apply. The PMPM cost of HEMS to a commercially insured population is approximately $0.65, suggesting that population-based funding mechanisms may be financially plausible under favorable reimbursement and utilization conditions. These findings can inform financial planning, payer negotiations, and policy design for air medical transport systems.

## 7. Acknowledgments

The authors wish to thank Dr. Sybil Russell for mentorship, support, and research guidance. Rabbi Yisroel Altein provided valuable ethical feedback on the research approach.

## 8. Funding Statement

Funding for this work was provided by The MITRE Corporation (ww.mitre.org) through Internal Research and Development (IR&D) funding. This was used to support the salaries of DA, DH, LRI, RDL, SA, SH, and VS. The sponsor played an advisory role in study design, data collection, and analysis. The decision to publish and the preparation of the manuscript were the sole responsibility of the authors.

## 9. Author Contributions

Conceptualization, RDL and SA; Methodology, RDL, SA; Formal analysis, LRI, DH; Data curation, LRI, DA, VS; Writing – original draft, RDL, SA; Writing – review & editing, all authors; Supervision, RDL, SA, SH.

## 10. Competing Interests

The authors declare no competing interests.

## 11. Data Availability Statement

Data and analysis tools supporting this study are publicly available at Zenodo (DOI: 10.5281/zenodo.20601704 and on the web at https://zenodo.org/records/20601705). Medicare fee schedule data are publicly available from CMS (https://hcpcs.codes/fee-schedule/ambulance/).

# S1 Appendix. Actuarial Rate Development Calculation Details

The calculation for each value in Table 3 is as follows:

- (a) is a utilization rate per 1,000 based on population characteristics, burden of disease, and availability (supply) = *Count* from Table 3 / 3,879,778 (Population) *1,000
- (b) is a rate set based on a provider's posted list price = *Sum* from Table 3 / *Count* from Table 3
- (c) = (a) * (b) / (12 months *1,000 lives)
- (d) is a feature of the individual insurance contract or program where patients pay a copayment or other portion of their bill; these amounts are notional
- (e) = (a) * [(b) – (d)] / 12 months *1,000 lives)